\documentclass[twocolumn,aps,prl,superscriptaddress,showpacs]{revtex4}

\usepackage{color}
\usepackage{graphicx}

\newcommand{\ket}[1]{\ensuremath{\left| #1 \right\rangle}}

\begin{document}

\title{Entanglement-enhanced measurement of a completely unknown phase}

\author{G.~Y. Xiang}
\affiliation{Centre for Quantum Computer Technology, Centre for Quantum Dynamics, Griffith University, Brisbane, 4111, Australia}
\author{B.~L. Higgins}
\affiliation{Centre for Quantum Computer Technology, Centre for Quantum Dynamics, Griffith University, Brisbane, 4111, Australia}
\author{D.~W. Berry}
\affiliation{Institute for Quantum Computing, University of Waterloo, Waterloo, ON N2L 3G1, Canada}
\author{H.~M. Wiseman}
\affiliation{Centre for Quantum Computer Technology, Centre for Quantum Dynamics, Griffith University, Brisbane, 4111, Australia}
\author{G.~J. Pryde}
\email{G.Pryde@griffith.edu.au}
\affiliation{Centre for Quantum Computer Technology, Centre for Quantum Dynamics, Griffith University, Brisbane, 4111, Australia}

\begin{abstract}
The high-precision interferometric measurement of an unknown phase is the basis for metrology in many areas of science and technology. Quantum entanglement provides an increase in sensitivity, but present techniques have only surpassed the limits of classical interferometry for the measurement of small variations about a known phase. Here we introduce a technique that combines entangled states with an adaptive algorithm to precisely estimate a completely unspecified phase, obtaining more information per photon that is possible classically. We use the technique to make the first \textit{ab initio} entanglement-enhanced optical phase measurement. This approach will enable rapid, precise determination of unknown phase shifts using interferometry.
\end{abstract}


\maketitle 

Precise interferometric measurement is vital to many scientific and technological applications. The use of quantum entanglement allows interferometric sensitivity that surpasses the standard quantum limit (SQL)~\cite{Giovannetti2004,WisMil2010}. Experimental demonstrations of entanglement-enhanced sub-SQL interferometry~\cite{Meyer2001,Leibfried2005,Nagata2007,Okamoto2008}, and most theoretical treatments~\cite{Caves1981,Yurke1986,SumPeg90,Holland1993,Sanders1995,Lee2002,Steuernagel2002,Hofmann2006,Cable2007,Dowling2008}, address the goal of obtaining an increased interference fringe gradient. This is suitable for sensing small variations about an already known phase, but does not give a self-contained measurement of an unknown phase anywhere in $[0,2\pi)$. Both tasks are important~\cite{WisMil2010}, but not equivalent, and to move from the phase-\emph{sensing} regime to the phase-\emph{measurement} regime requires one of several nontrivial measurement algorithms~\cite{Higgins2009,Berry2009}. Here, we demonstrate the first sub-SQL \emph{measurement} of an unknown phase using entanglement-enhanced optical interferometry. Our technique uses a ``bottom-up'' approach, making optimal use of whatever (typically imperfect) entanglement is available to obtain the phase estimate most efficiently.

Obtaining phase sensitivity by using entanglement yields an in-principle advantage in bandwidth over recent demonstrations of sub-SQL phase measurement using sequences of multiple passes of single photons~\cite{Higgins2007,Higgins2009}. Although such techniques avoid the complexities of generating entangled states, and are suitable for measuring static phase shifts, they are unsuitable for fast measurement because the time $t$ to complete a measurement scales as the total number of photon passes $N$. Applications like the measurement of rapidly varying phase shifts, or rapid measurement of multiple samples, require a technique where increasing precision does not significantly decrease bandwidth. This can only be achieved by entangled states.

A suitable technique for achieving sub-SQL phase measurement using entangled states is to apply the measurement algorithm of Ref.~\cite{Higgins2007} to a sequence of entangled $n$-photon ``NOON'' states~\cite{Dowling2008,Walther2004,Mitchell2004}, which have optimal phase sensitivity for a given $n$. In this case the measurement time $t$ scales as $\log N$, as opposed to $N$ for the multipass implementation. NOON states, however, are notoriously difficult to generate, even for moderate $n$. Previous investigations into exploiting entanglement-enhanced sensitivity have employed a ``top-down'' approach, starting with a theoretical knowledge of the optimal states and determining how to approximate these experimentally by constructing complex circuits to filter them from more easily produced states, and using only some measurement results. By contrast, we adopt a ``bottom-up'' approach by taking available entangled states and using all measurement results to obtain the most phase information.
Our scheme uses Bayesian analysis and optimized adaptive feedback~\cite{BerWis00,Higgins2007}. In contrast to the algorithm of Ref.~\cite{Higgins2007}, we use a general approach that can be applied to any entangled state, including NOON states (should efficient production become available in the future).

\begin{figure}[!tbh]
\center{\includegraphics[width=4cm]{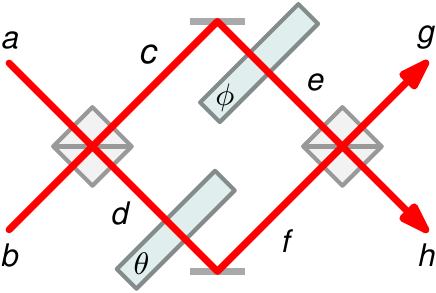}}
\caption{\label{fig:interferometer} A Mach-Zehnder interferometer for interferometric phase measurement. Photon states in modes $a$ and $b$ are incident on a beam splitter. In general, nonclassical interference generates entanglement in the output modes of the beam splitter, $c$ and $d$. After passing though the phase shifts $\phi$ (unknown) and $\theta$ (controllable), giving modes $e$ and $f$ respectively, the modes recombine on the final beam splitter, and the interferometer output modes $g$ and $h$ propagate to detectors. With knowledge of the input states, $\theta$, and the detection result, knowledge is gained about the phase shift $\phi$.}
\end{figure}

In this experiment, we consider the states produced by $n$-photon dual Fock state inputs, i.e.\ states of the form $\ket{n/2,n/2}_{a,b}$, to the first beam splitter of an interferometer, as shown in Fig.~\ref{fig:interferometer}. These states have been shown to be capable of phase sensing at the Heisenberg limit, that is, with Fisher length scaling as $1/N$ \cite{Holland1993,Berry2001,Okamoto2008}. We generate these states by spontaneous parametric down-conversion (SPDC) and post-selection on counting a given total number of photons. However any source of pairs of indistinguishable single-mode $n$-photon states could be employed. In our experiment, we use both the 2-photon state $\ket{1,1}$ and the 4-photon state $\ket{2,2}$ (as well as the single-photon input $\ket{1,0}$) at different stages of the measurement protocol.

Nonclassical interference of the dual Fock states at the first beam splitter produces photon number entanglement in the two arms of the interferometer, $c$ and $d$. For a $\ket{1,1}_{a,b}$ input, the state inside the interferometer is $(\ket{2,0}_{c,d} + \ket{0,2}_{c,d})/\sqrt{2}$.  This is an $n=2$ NOON state, and corresponds to the well-known Hong-Ou-Mandel effect~\cite{Hong1987}. With the unknown random phase shift $\phi$ in one arm of the interferometer, and a controllable ``feedback'' phase shift $\theta$ in the other arm, this state evolves to $(e^{2i\phi}\ket{2,0}_{e,f} + e^{2i\theta}\ket{0,2}_{e,f})/\sqrt{2}$. The phase factor $e^{2i\phi}$ demonstrates phase \emph{super-resolution} \cite{Eisenberg2005,Sun2006,Resch2007}---in contrast to a single photon input $\ket{1,0}_{a,b}$, for which the state acquires  a phase factor of only $e^{i\phi}$. Generally, $n$-photon NOON states exhibit $n$-fold super-resolution, and it is this super-resolution that gives such states (in principle) the best possible phase sensitivity.

While nonclassical interference acting on an $\ket{n/2,n/2}_{a,b}$ state continues to generate entangled states as $n$ increases, these states are not NOON states for $n > 2$. The 4-photon $\ket{2,2}_{a,b}$ input results in a state inside the interferometer of $\sqrt{3/8} (\ket{4,0}_{c,d} + \ket{0,4}_{c,d}) - \ket{2,2}_{c,d}/2$,  which evolves to $\sqrt{3/8} (e^{4i\phi}\ket{4,0}_{e,f} + e^{4i\theta}\ket{0,4}_{e,f}) - e^{2i(\phi+\theta)}\ket{2,2}_{e,f}/2$. While this state is entangled, and exhibits components  with a 4-fold increase in phase resolution, it also contains an extra term ($\ket{2,2}_{e,f}$) with only a 2-fold increase.

Interestingly, it was theoretically shown~\cite{Steuernagel2002} and subsequently experimentally demonstrated~\cite{Nagata2007} that NOON-like 4-photon phase super-resolution can be extracted from the state generated by the $n=4$ dual Fock input if post-selection is employed. Refs.~\cite{Nagata2007} and \cite{Okamoto2008}---the latter using an improved experiment and more thorough analysis---show that phase \emph{sensing} below the SQL is possible by this method, even taking into account the discarding of certain results.
However, it is obviously not optimal to deliberately throw away phase information. Here, we do not select a subset of output components---instead, we use the full phase information encoded in the state.

Our scheme is as follows. We employ a sequence of entangled states---a phase shift $\phi$ is measured using $M_k$ instances of each entangled $n_k$-photon state, where $n_k=2^k$ and $k \in \{0,1,2\}$. We begin with a flat phase probability distribution $P(\phi)=1/2\pi$, but after each measurement is performed, knowledge about the phase is updated by applying Bayes' theorem to $P(\phi)$. The feedback phase $\theta$ is initially random, but after a detection it is always set to minimize the \emph{expected} phase variance after the subsequent detection, following the algorithm of Ref.~\cite{BerWis00}. The total resources used are quantified by the total photon number, $N=\sum_k 2^k M_k$. We perform an exhaustive numerical search to determine the optimal (or near optimal) $M_k$ and $n_k$ for a given $N$.

We examine the expected behaviour of an ideal implementation for the various $n$-photon inputs. Single photons ($n_k = 1$) incident on the first beam splitter of the interferometer are sufficient to generate the (trivial) $n=1$ NOON state. The photon number difference $\Delta$ between the two outputs of the final beamsplitter of the interferometer can take two possible values, $\Delta = \pm 1$, which we rewrite as $\Delta = -1+2x$, where $x\in\{0,1\}$. The probabilities for these two outcomes are 
\begin{equation}
P_1 \left( \Delta = -1 + 2x \, | \, \phi, \theta \right) = A_{x,0} + A_{x,1} \cos \left( \phi - \theta \right), \label{eq:P1}
\end{equation}
where the $2\times2$-matrix $A$ is defined as
\begin{equation} 
A = \frac{1}{2}\left[\matrix{1 & 1 \cr 1 & -1 \cr}\right].
\end{equation}

For the $\ket{1,1}_{a,b}$ input ($n_k = 2$), which produces a 2-photon NOON state inside the interferometer, the probabilities for photon detection at the outputs are
\begin{equation}
P_2 \left( \left| \Delta \right| = 2x \, | \, \phi, \theta \right) = B_{x,0} + B_{x,1} \cos \left[ 2 \left( \phi - \theta \right) \right]
\end{equation}
where the matrix $B=A$ in this ideal case, and $x\in\{0,1\}$ as above. (In general the sign of $\Delta$ matters only for odd $n$.)

For the $\ket{2,2}_{a,b}$ input ($n_k = 4$), the probability for each combination of number states at the outputs of the interferometer can be written as
\begin{equation}
P_4\left( \left| \Delta \right| = 2x \, | \, \phi, \theta \right) = \sum_{y=0}^2 \Gamma_{x,y} \cos \left[ 2y \left( \phi - \theta \right) \right]
\end{equation}
where $x \in \{0,1,2\}$ and
\begin{equation}
\Gamma = \frac{1}{32} \left[\matrix{11 & 12 & 9 \cr 12 & 0 & -12 \cr 9 & -12 & 3 \cr}\right]. \label{eq:Gamma}
\end{equation}
Equations~(\ref{eq:P1})--(\ref{eq:Gamma}) define the probability functions that allow us to construct the Bayesian updating protocol.

\begin{figure}[!tbh]
\center{\includegraphics[width=7cm]{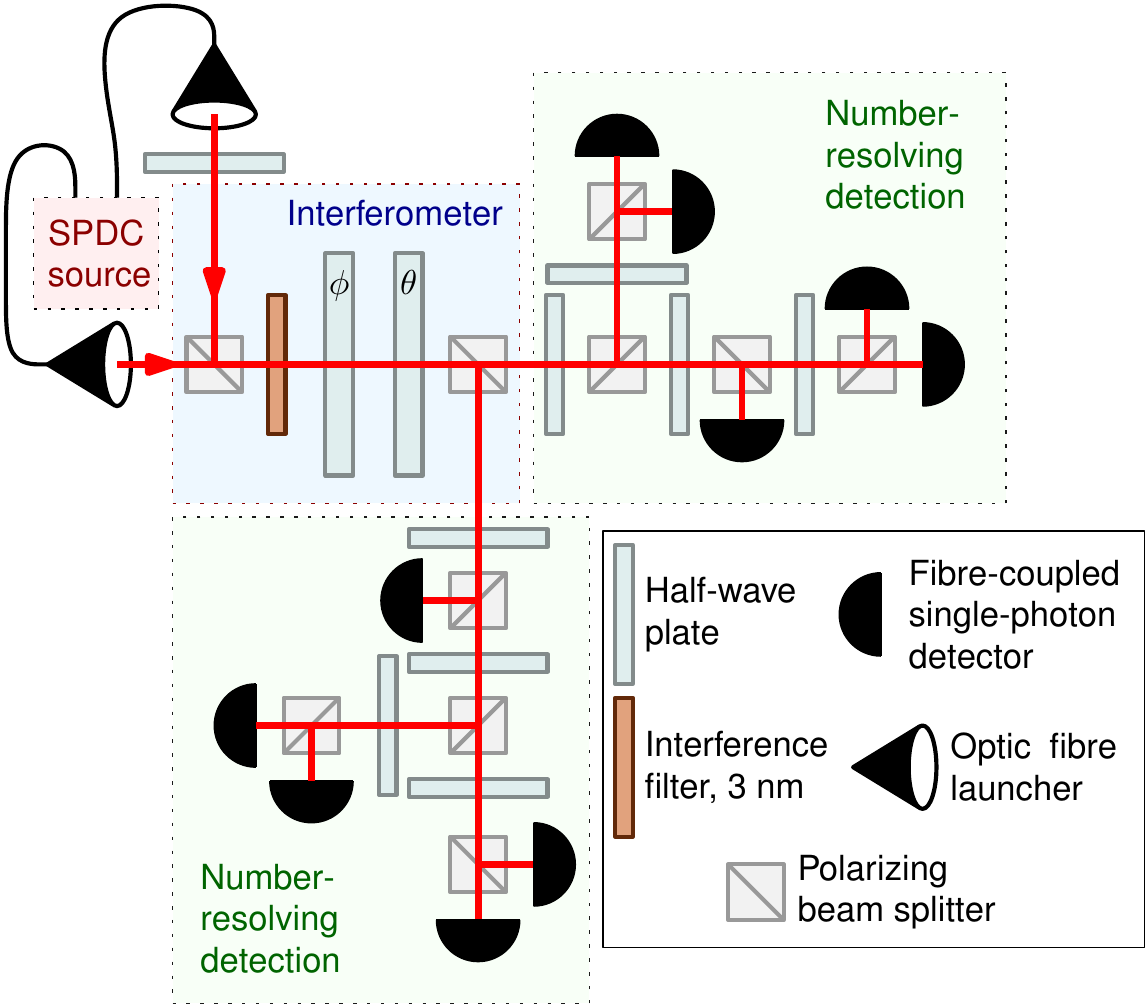}}
\caption{\label{fig:layout}Layout of the experiment. A spontaneous parametric down-conversion (SPDC) source produces pairs of photons and biphotons which implement our dual Fock states, and are guided to the interferometer with single-mode polarization maintaining optic fibres. These photon states are incident on a polarizing beam splitter and undergo phase shifts in the left- and right-circular polarization modes due to the $\phi$ and $\theta$ half-wave plates. The final beam splitter recombines the two interferometric modes, and the output states are measured in the photon number basis by single-photon counting module (SPCM) arrays. The results of all previous measurements determine $\theta$ for the next measurement. For single photon inputs, one optic fibre from the SPDC source is redirected to a single-photon counting module, with single-photon counting performed in coincidence.}
\end{figure}

Our experimental demonstration uses a common-spatial-mode polarization interferometer, as in Fig.~\ref{fig:layout}. A type-I BBO crystal is pumped by a frequency-doubled mode-locked Ti:Sapphire laser and coupled to polarization-maintaining optical fibres. The resulting spontaneous parametric down-conversion supplies the interferometer with pairs of 820~nm single photons and pairs of biphotons. One horizontally polarized mode and one vertically polarized mode are combined into a single spatial mode using a polarizing beam splitter. The right- and left-circular polarization modes of this single spatial mode constitute the arms of the interferometer, and contain the 2- and 4-photon entangled states. Phase shifts between these circular polarizations are performed using half-wave plates, implementing the unknown $\phi$ and controllable $\theta$ phases. We implement photon number detection at the outputs of the interferometer by evenly splitting each beam into an array of single-photon detectors. For measurements with single photons, one output arm of the SPDC source is guided directly to a detector, and the single photon is heralded by detection coincident with that detector.

While theoretical analyses typically assume idealized states, imperfections in the experimental apparatus lead to non-idealities in the real photon states that are generated. To demonstrate the full power of our approach, we include knowledge of these non-idealities in our Bayesian updating mechanism. Obtaining this knowledge requires careful characterization of our apparatus, which we do by least-squares fits to phase fringe data collected with the system phase $\phi$ absent. From this (see Appendix A for details) we arrive at the experimental detection probability coefficient matrices:
$$
A' = \frac{1}{2} \left[\matrix{0.999 & 0.976 \cr 1.001 & -0.976 \cr}\right] \quad
B' = \frac{1}{2} \left[\matrix{ 0.989 & 0.940 \cr 1.011 & -0.940 \cr}\right]
$$
\begin{equation}
\Gamma' = \frac{1}{32} \left[\matrix{ 11.206 & 9.829 & 7.596 \cr 12.901 & 0.595 & -10.192 \cr 7.893 & 10.423 & 2.596 \cr }\right]. \label{eq:ExptChar}
\end{equation}
It is these experimentally determined coefficients which we use to determine the optimal sequences of input configurations $(n_k, M_k)$.   For example, we find that for $N=37$ resources the optimal sequence adaptively measures eight biphotons, followed by nine single photons, and finally three 4-photon states. We experimentally demonstrate our algorithm for a representative sample set of $N\in \{4,9,15,25,37,48\}$ resources---the full set of $(n_k, M_k)$ configurations used for these $N$ can be found in Appendix B.

\begin{figure}[!tbh]
\center{\includegraphics[width=8cm]{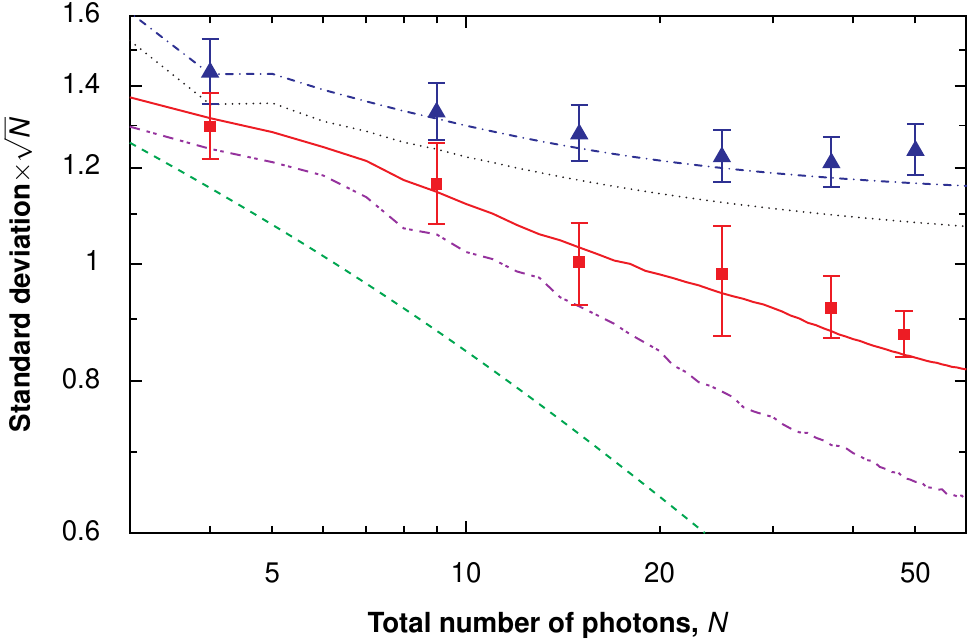}}
\caption{\label{fig:data}Standard deviations of phase measurement for varying total photon number $N$. A multiplying factor of $\sqrt N$ results in the SQL asymptoting to a horizontal line. The red solid line and red boxes show the theoretical and experimentally measured performance, respectively, of our phase measurement scheme. This clearly surpasses the SQL (black dotted line) and the results of an SQL scheme demonstrated using the same apparatus (blue dash-dotted line, theory; blue triangles, experiment). Theoretical performance of the scheme given unit visibility states and optimal sequences is shown as the purple dash-dot-dot line. The green dashed line represents the fundamental limit due to Heisenberg's uncertainty principle.}
\end{figure}

The results of our phase measurements are plotted in Fig.~\ref{fig:data}, together with theoretical predictions. Each data point represents 1000 estimates, with error bars showing 95\% confidence intervals calculated using a bootstrap sampling method with $\approx10^6$ samples~\cite{Davison1997}. For comparison, we have also demonstrated a standard-quantum-limited phase measurement scheme with the same experimental layout. This scheme uses only single-photon states, with the (initially random) controllable phase $\theta$ incremented nonadaptively by $\pi/N$ after each measurement. With perfect single-photon visibility this scheme defines the  SQL; the imperfect single-photon visibility of the experiment ($\approx 97.6\%$)  means that our experimental implementation yields phase uncertainties above the SQL, in agreement with theoretical predictions.

The adaptive scheme clearly operates with phase uncertainty below the SQL,  with an improvement that increases monotonically with total photon number $N$. For  the highest $N$ demonstrated (48) we experimentally measure uncertainty more than 1.86~dB better than the theoretical SQL, and more than 2.85~dB better than the corresponding measured uncertainty of the single-photon scheme for this interferometer visbility. The unknown phase $\phi$ is fixed, but our choice of an initially random $\theta$ ensures equivalence to measuring an unknown phase $\phi \in [0,2\pi)$. Thus these results show, for the first time, the demonstration of a scheme which beats the SQL for the measurement of a random phase using entangled states. 

It is important to note that, in principle, our bottom-up approach can consider not just enangled states produced from $\ket{n/2,n/2}$ inputs, but any photon number state inside the interferometer, including ideal $n$-photon NOON states. Given such NOON states with $n_k \in \{1,...,2^K\}$, the optimization procedure used here will find a sequence that performs at least as well as the sequence of Ref.~\cite{Higgins2007}, which achieved scaling of phase uncertainty at the fundamental limit of precision due to Heisenberg's uncertainty principle. Other phase sensitive states~\cite{Chiruvelli2009}, including loss resistant states \cite{Kacprowicz2009,Dorner2009,Lee2009}, might also be considered using our approach.
 
We have proposed and demonstrated a powerful and  general bottom-up approach to the measurement of random optical phase $\phi \in [0, 2\pi)$, employing Bayesian analysis and optimal adaptive feedback to make the best use of available photon states. This is the first demonstration of sub-SQL measurement of a random phase using entangled states, which can potentially achieve high bandwidth in quantum-enhanced phase measurements, with a wide range of metrological applications.

\acknowledgments

We thank Jeremy O'Brien for helpful discussions. This work was supported by the Australian Research Council.

\appendix
\appendix
\section{Appendix A: Photon number detection}
We use two five-detector arrays of single-photon counting modules to implement number-resolved detection. Each output mode $e$ and $f$ is split into 5 separate spatial modes, one for each single-photon detector, using half-wave plates and polarising beam splitters. The half-wave plates are set such that an equal proportion of an output mode is incident on each photon detector for that mode. The layout of the single-photon detectors is asymmetric for logistical reasons.

$n$-photon states ($n>1$) are signaled by coincident detection of $n = n_e + n_f$ photons across the detectors, where $n_e$ and $n_f$ represent the number of photons detected in the respective output modes. With 5 detectors in each of the two output arms of the interferometer, there are a total of $^{10}C_n$ possible coincidence detection patterns that describe an $n$-photon output state---for 4-photon states this gives 210 patterns.

The projection probability, that is, the probability that a particular photon output state $\ket{n_e, n_f}_{e,f}$ will be successfully resolved, depends on $n_e$ and $n_f$ even if the individual detectors are unit-efficiency (but not photon-number resolving) photodetectors. For example, in this unit-efficiency case the 4-photon $\ket{2,2}_{e,f}$ state has a projection probability of 0.64 with this detection scheme, whereas the $\ket{4,0}_{e,f}$ state has a projection probability of only 0.096. Like many other experiments, we do not consider loss in our calculation of resources $N$. However, we require that the probability of projection is independent of the particular output state of the interferometer.

In addition, a technical limitation means that we can only consider a maximum of 128 patterns at once. For these reasons, we consider only a limited set of patterns, randomly chosen for each measurement result, such that the ultimate probability of detection for each result is approximately independent of the state. We use as many patterns as we can up to the limit of our electronics. To address the remaining discrepancy, we randomly discard a certain small proportion of measurement results in software, before the result can be used in the algorithm. This is equivalent to introducing a controlled state-dependent loss. We emphasize that this solution is a consequence of the imperfect number detection mechanism we use, necessary to simulate perfect detectors, and is not fundamental to our approach.

We determine the appropriate proportion of introduced state-dependent loss from our phase fringe characterization of the experiment, which is done with the system phase $\phi$ absent, and given the limited set of detection patterns. From least squares fits to the count rates obtained with $\theta$ varied over the range $[-\pi, \pi]$, we derive three matrices $J$ similar to those of Eq.~\ref{eq:ExptChar}. By taking the first column of the inverse of each matrix we obtain the state-dependent loss probabilities:
\begin{center}
\begin{tabular}{|c|c|}
\hline
Detected State & Loss Probability \\
\hline
\hline
$\ket{1,0}_{e,f}$ & 0 \\
$\ket{0,1}_{e,f}$ & 0.1276 \\
\hline
$\ket{1,1}_{e,f}$ & 0.1975 \\
$\ket{2,0}_{e,f}$ or $\ket{0,2}_{e,f}$ & 0 \\
\hline
$\ket{2,2}_{e,f}$ & 0.2304 \\
$\ket{3,1}_{e,f}$ or $\ket{1,3}_{e,f}$ & 0.3395 \\
$\ket{4,0}_{e,f}$ or $\ket{0,4}_{e,f}$ & 0 \\
\hline
\end{tabular}
\end{center}
Doing so also ensures the detection probability is independent of the phase, which is a necessary condition of the Bayesian algorithm. We can then apply these probabilities to the fit parameter matrices $J$ to obtain the values of Eq.~\ref{eq:ExptChar}.

\section{Appendix B: Sequence configurations}
Our approach determines the optimal sequence of $M_k$-many $n_k$-photon states using an exhaustive numerical search. The sequences we demonstrate are:
\begin{center}
\begin{tabular}{|c||c||c|c||c|c||c|c|c||c|c|c||c|c|c|}
\hline
$N$ & 4 & \multicolumn{2}{|c||}{9} & \multicolumn{2}{|c||}{15} & \multicolumn{3}{|c||}{25} & \multicolumn{3}{|c||}{37} & \multicolumn{3}{|c|}{48} \\
\hline
\hline
$M_k$ & 4 & 7 & 1 & 9 & 3 & 13 & 4 & 1 & 8 & 9 & 3 & 10 & 8 & 5 \\
\hline
$n_k$ & 1 & 1 & 2 & 1 & 2 & 1 & 2 & 4 & 2 & 1 & 4 & 2 & 1 & 4 \\
\hline
\end{tabular}
\end{center}
Note that, as our scheme is adaptive, the left-to-right ordering of sequences is significant.

\section{Appendix C: Fisher information} 
The Fisher information generated by a phase-sensitive measurement is defined by
\begin{equation}
F(\phi) = \sum_{x} \frac{1}{P(x|\phi)} \left(\frac{\partial P(x|\phi)}{\partial\phi}\right)^2,
\end{equation}
where $P(x|\phi)$ is the probability of measurement result $x$ given that the true system phase is $\phi$. The Fisher information places a lower bound on the smallest possible \emph{shift} $\delta \phi$ in the phase away from $\phi$ that can be reliably detected from a large number $M$ of repeated measurements, via 
the Cram\'{e}r-Rao inequality, 
\begin{equation}
\delta \phi \le 1/\sqrt{M \times F(\phi)}.
\end{equation} 
This motivates defining the Fisher length as $1/\sqrt{F}$. For ideal measurements on an $N$-photon NOON state the Fisher information is $N^2$ and Fisher length is $1/N$, independent of the system phase. Thus the Fisher information for a 4-photon NOON state, for example, is $16$. It is additive for independent measurements on two separate states, so the Fisher information for two 2-photon NOON states is $8$, half that for a single 4-photon NOON state.

\begin{figure}[!tbh]
\center{\includegraphics[width=8cm]{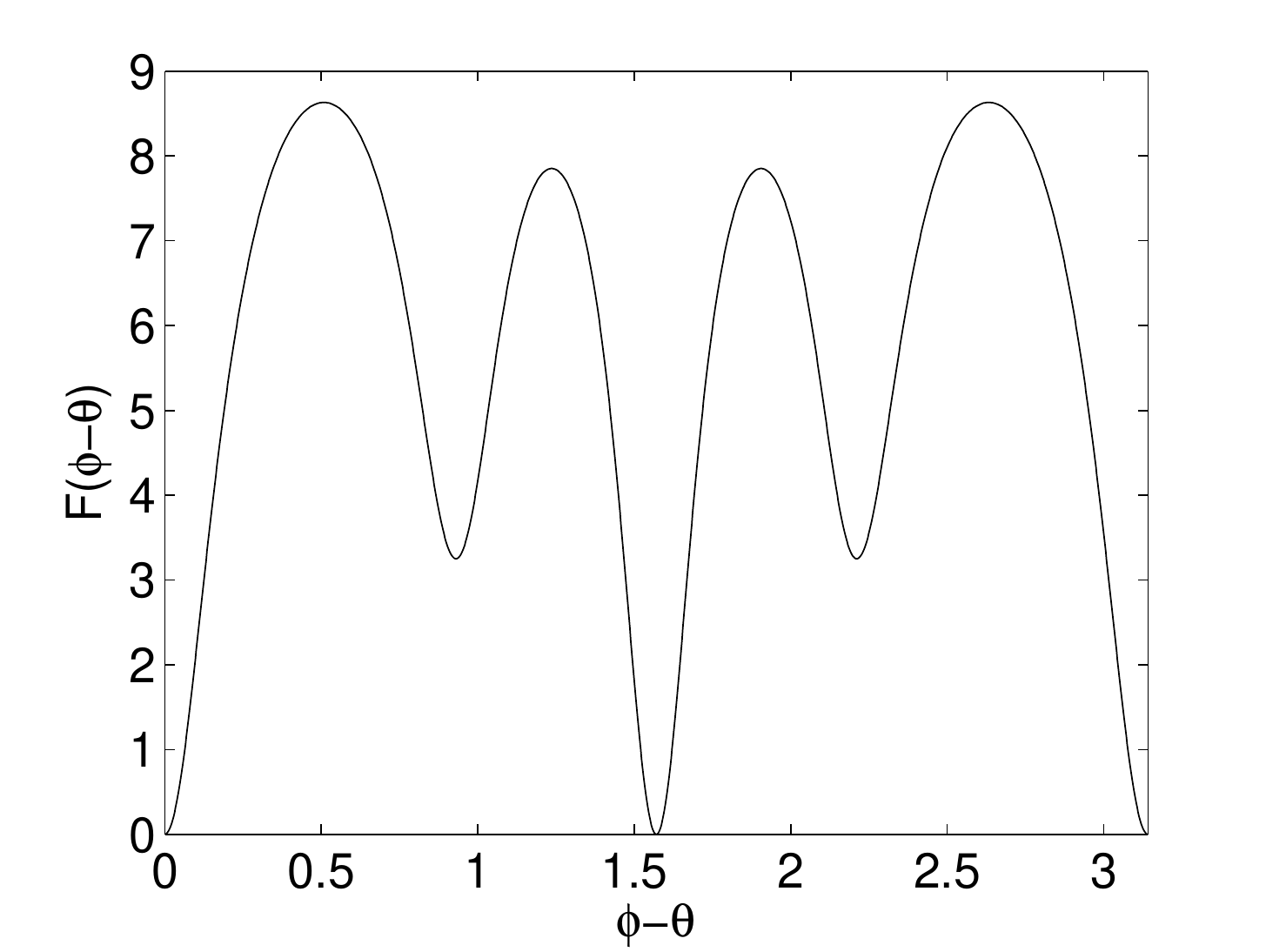}}
\caption{\label{fig:fisher} The Fisher information for measurements on the $\ket{2,2}_{a,b}$ input state with the experimental detection probability matrix $\Gamma'$ given by Eq.~\ref{eq:ExptChar}.}
\end{figure}

For the 4-photon states that we generate, the Fisher information is less than that of a 4-photon NOON state, and is equal to $12$ for ideal measurements. With the experimental detection probability matrix $\Gamma'$ given in Eq.~\ref{eq:ExptChar} for our 4-photon input state $\ket{2,2}_{a,b}$, the maximum Fisher information is $8.6$. This is above the value of $8$ for two independent 2-photon ideal NOON states, and well above the value for two 2-photon NOON states with our experimentally measured visibilities, which is at most $7.1$.

With the experimental $\Gamma'$ matrix, the Fisher information is no longer independent of $\phi$, and has the dependence shown in Fig.~\ref{fig:fisher}. With the addition of the controllable phase $\theta$, the Fisher information is a function of $\phi-\theta$. The sensitive dependence on the system phase is likely to be the reason why it is often optimal to perform the 4-photon measurements last---the system phase must already be known quite accurately in order to adjust the feedback phase to maximise the Fisher information.

\end{document}